\documentclass[11pt]{article}
\usepackage{cite,moriond,epsfig}
\usepackage[latin1]{inputenc}
\usepackage[OT1]{fontenc}
\usepackage{amssymb,amsmath}

\newcommand{\LT}{\left[}
\newcommand{\RT}{\right]}
\newcommand{\LF}{\left(}
\newcommand{\RF}{\right)}

\def\be{\begin{equation}}
\def\ee{\end{equation}}
\def\bea{\begin{eqnarray}}
\def\eea{\end{eqnarray}}

\def\mycite#1{[\citen{#1}]}
\begin{document}
\vspace*{0cm}
\title{Features and nongaussianity\\ in the inflationary
power spectrum}

\author{James M.\ Cline }

\address{Department of Physics, McGill University, 3600 University
St., Montr\'eal, Qc H3A2T8, Canada}

\maketitle\abstracts{I summarize recent work on (1) constraining 
spike-like features in the cosmic microwave background (CMB) and large scale
structure (LSS); (2) nonstandard Friedmann equation in stabilized warped 6D
brane cosmology, with applications to inflation; and (3) nonlocal
inflation models, motivated by string theory, which can yield large
nongaussian CMB fluctuations.  Work in collaboration with 
N.\ Barnaby, T.\ Biswas, F.\ Chen, L.\ Hoi, G.\ Holder and S.\ Kanno. }

\section{Features in the CMB and LSS}

It is hoped that with improvements in the CMB and LSS data, 
fingerprints of the underlying inflationary model may be discovered,
in the form of features going beyond the amplitude and tilt of the
spectrum.  One such feature is the appearance of bumps in the
spectrum, which can be caused for example by a tachyonic instability
during inflation.\cite{JL}\ In ref.\ \mycite{Hoi:2007sf} we undertook a
systematic search for evidence of such bumps using the latest CMB and
LSS data in conjunction with a modified CosmoMC code.   Our initial
ansatz for the shape of the bump was motivated by a number of models
\cite{JL,BC,Gong} and theoretical arguments \cite{Traschen} for a
scale noninvariant addition to the spectrum of the form $\delta P = A
k^3$ (thus having spectral index $n_\sigma=4$), cut off at some maximum
wave number $k_c$. This is a model with two extra parameters, $A$ and
$k_c$, similar to the running spectral index extension of the
standard $\Lambda$CDM model, since the latter also introduces the
tensor ratio $r$.  As we will discuss, the results are rather
insensitive to the value $n_\sigma=4$, so it makes sense to fix this
exponent based on theoretical expectations and keep only two extra
parameters.  

\begin{figure}[h]
\centerline{\psfig{figure=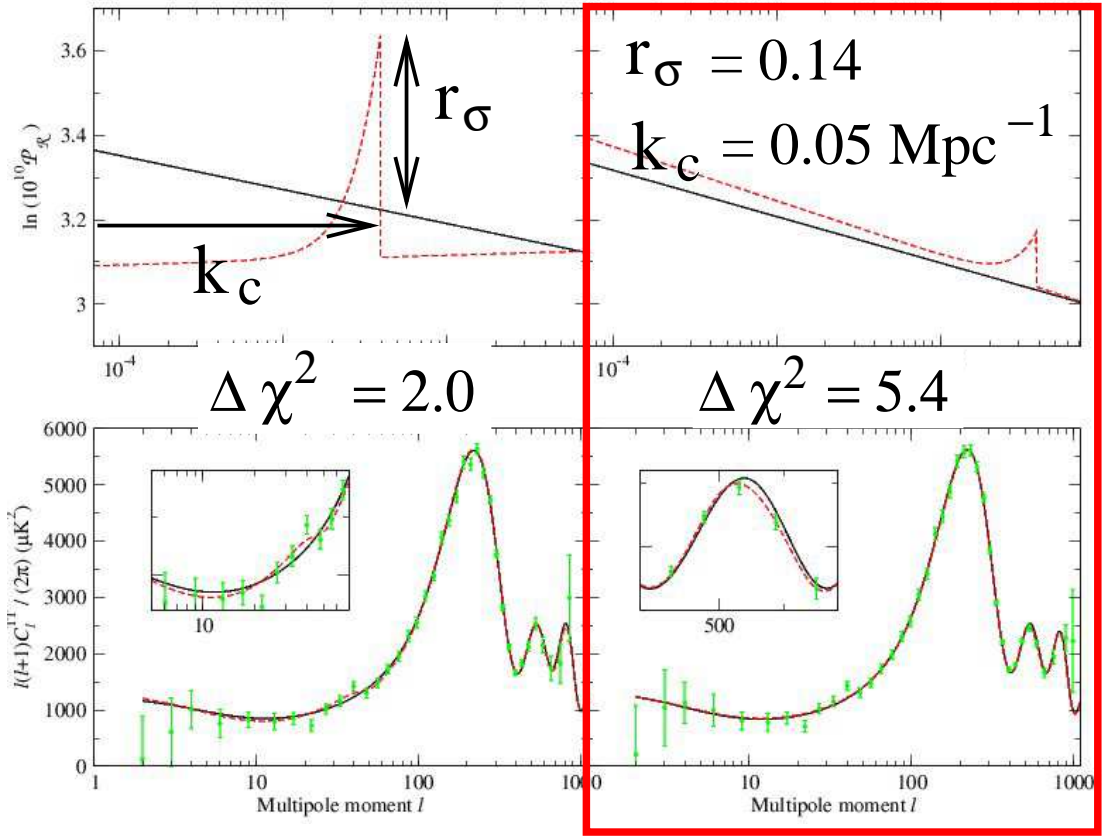,height=4.4in}}
\caption{Top panels: standard $\Lambda$CDM power spectrum (solid
lines) and best-fit spectrum perturbed by spike (dashed line).
Bottom panels: corresponding temperature (TT) multipoles,
showing improvements in fit to data.
\label{fig1}}
\end{figure}

In figure \ref{fig1} the top panels show two examples of best-fit 
cut-off $k^3$ spikes found by CosmoMC, superimposed on the best fit
$\Lambda$CDM spectrum.  We trade the absolute amplitude $A$ of the
spike for the parameter $r_\sigma$ which denotes its ratio to the
amplitude of the standard, nearly-scale-invariant component.  The
bottom panels show how these spikes in the primordial spectrum,
combined with the adjustments of the other cosmological parameters,
improve the fit of the model to the WMAP3 data in the two regions,
where the standard $\Lambda$CDM prediction (solid curves) either fell
above or below the data.  The improvement to the fit is most
significant for the second spike, in the vicinity of multipole $\ell
= 550$,  giving a reduction in $\chi^2$ of 5.4, notably more
significant than the improvement which was obtained by the
much-discussed running spectral index model.   The combination of 
the spike with a smaller value of $n_s$ for the scale invariant
component allows for the prediction to be lowered in the vicinity of
the feature, and thereby to better match the data, relative to the $\Lambda$CDM
prediction. It is striking that such large spikes, and accompanying
large changes in $n_s$ for the scale invariant component, are not
only allowed by the data, but even preferred by it.  

Since finishing ref.\ \mycite{Hoi:2007sf}, the WMAP5 data have been
released and we have rerun the code with the new
data.\footnote{Referees of JCAP have delayed the publication of ref.\
\mycite{Hoi:2007sf} even though it was completed significantly before
WMAP5.  The objection of the referees was entirely based on a
dislike of the $k^3$ ansatz, rather than any qualms with our analysis
of the data.}  Interestingly, just like the evidence for the running
spectral index, the evidence for the spikes disappears using WMAP5.
Instead, we get upper bounds on the amplitude ratio $r_\sigma$ as a
function of the cutoff $k_c$.  These will be published in an updated
version of ref.\  \mycite{Hoi:2007sf}.

One may question the assumption of the spectral index $n_\sigma=4$
for the nonscale-invariant perturbation.  Different values would give
rise to a broader or narrower spike.  However, these differences in
the primordial spectrum tend to get washed out in the multipoles,
which involve a convolution of the primordial power.  Table 1
gives the change in $\chi^2$ for the two features shown in figure
\ref{fig1} when $n_\sigma$ varies between 1 and 5; the variation in
the improvement to the fits is largely insignificant.

\begin{table}[h]
\begin{center}
\begin{tabular}{|c|c|c|}
\hline
$k^n$ & $\Delta\chi^2,\ \ell\sim 50$  & $\Delta\chi^2,\ \ell\sim 540$\\
\hline
$k^1$  &-0.1 &   4.1\\
$k^2$ &1.2 &   4.1\\
\underline{$k^3$} &  \underline{2.0} &   \underline{5.4}\\
$k^4$ &1.5 &   4.8\\
$k^5$ &1.9 &   5.4\\
\hline
\end{tabular} \end{center}
\caption{Comparison of fits using different spectal index for the
nonscale-invariant perturbation.}
\end{table}

\newpage

\section{Modified Friedmann equation and inflation from a 6D
braneworld model} 
\begin{figure}[h]
\centerline{\includegraphics[width=0.6\textwidth]{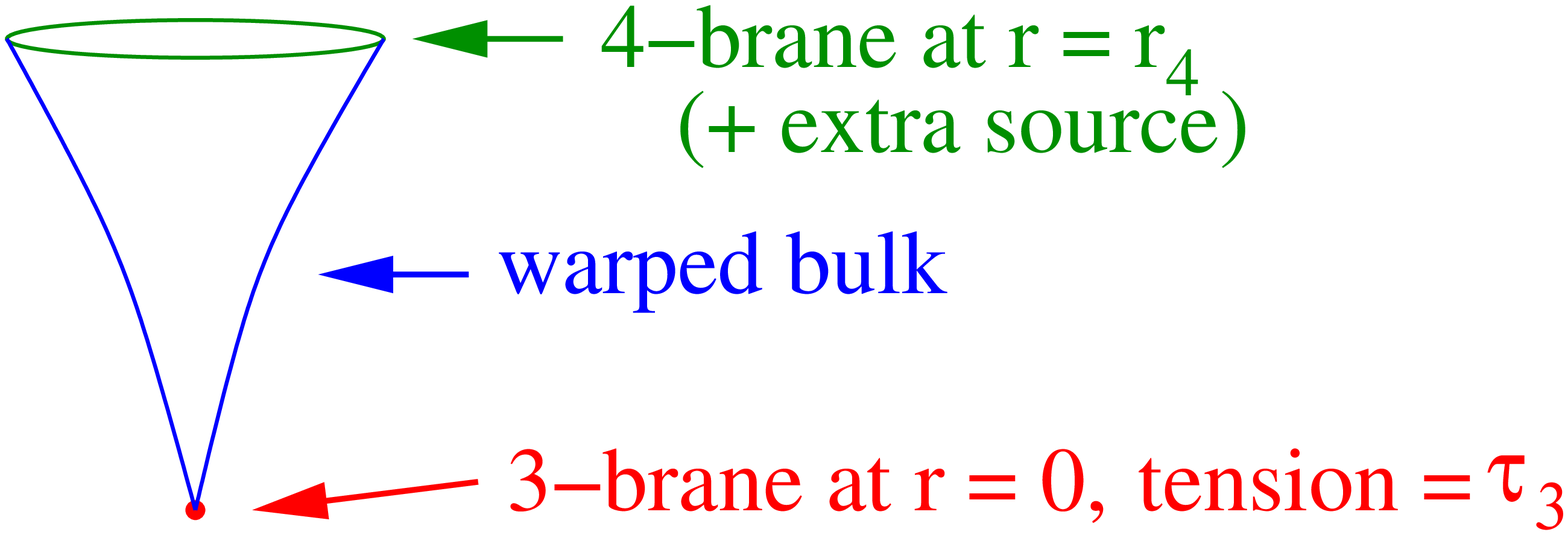}}
\caption{A 6D version of the Randall-Sundrum model, the AdS soliton
geometry.
\label{fig2}}
\end{figure}

In ref.\ \mycite{Chen} we have studied the closest
generalization of the Randall-Sundrum I model \cite{RSI} from 5 to 6 dimensions. 
The geometry is pictured in fig.\ \ref{fig2}: it is a warped (nearly
AdS) throat with a conical singularity (3-brane) at the bottom, and
cut off by a 4-brane at the top, with $Z_2$ orbifold boundary
conditions.  The cosmology of this model was studied earlier in 
ref.\ \mycite{CDGV}, which claimed that the standard Friedmann equation
was not recovered even when the extra dimensions were stabilized
using a bulk scalar field.  In ref.\ \mycite{Chen} we have corrected
this claim, demonstrating that general relativity (GR) is indeed
the valid description at low energies.

\begin{figure}[h]
\centerline{\includegraphics[width=0.6\hsize,angle=0]{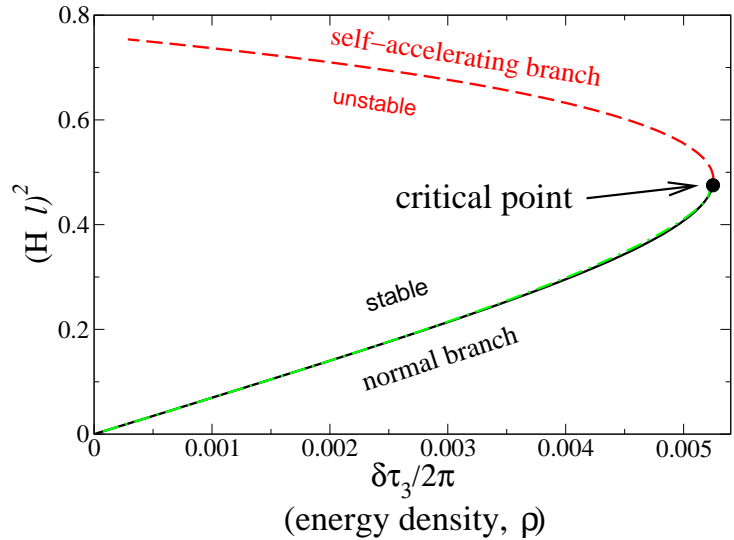}}
\caption{Hubble rate (squared) versus excess brane tension in 
6D model.
\label{fig3}}
\end{figure}

However there are interesting deviations to the Friedmann equations
at high scales, illustrated in fig.\ \ref{fig3}. For given values of
the 4-brane tension and bulk cosmological constant, one can find a
value of the 3-brane tension for which the 4D vacuum energy vanishes
and the solution is static.  We model the energy density of the
visible universe as excess tension $\delta\tau_3$ which gives rise to
expansion.   The Hubble rate is double valued as a function of
$\delta\tau_3$, and the two branches smoothly join each other (the
``critical point'' in fig.\ \ref{fig3}) at some maximum value of the
tension, $\delta\tau_{c}$.  The existence of two possible solutions
for the same sources of stress-energy also occurs in the DGP
model,\cite{DGP} where the self-accelerating branch is known to be
afflicted with a ghost.  In our case, the analysis of small
fluctuations reveals that the sickness of the exotic branch is
associated with a tachyonic instability of the radion, {\it i.e.},
the size of the extra dimensions.  Under small perturbations, the
solutions associated with the exotic branch decompactify.   Therefore
we disregard the high-$H$ branch in the following.

\begin{figure}[h]
\centerline{\includegraphics[width=0.6\hsize,angle=0]{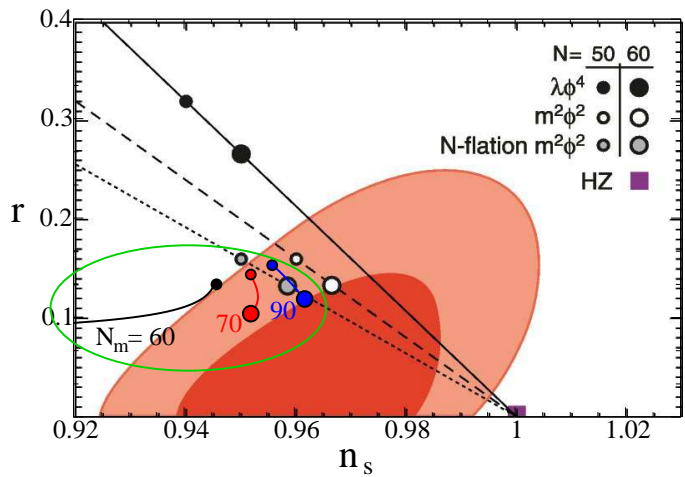}}
\caption{Predictions of chaotic inflation (and $N$-flation) and WMAP5
constraints for tensor ratio versus scalar spectral index.  The
predictions for the modified Friedmann equation are within the oval,
for $N_m$ (the maximum number of $e$-foldings due to the energy
density being bounded above by $\rho_c$) $=60$, 70 and 90.  Each curve
shows the evolution of the predicted $r,n_s$ as a function of number
of e-foldings after horizon crossing, $N$, for $N=50$ to 60.
Figure
adapted from ref.\ \protect\mycite{Spergel:2006hy}.\hfill
\label{fig4}}
\end{figure}

Nevertheless, interesting departures from GR occur near the critical
energy density, $\rho_c$.  They can be parametrized through a function of the
energy density on the 3-brane, ${\cal F}(\rho)= 1 + O(\rho/\rho_c)^2$,  by
writing the Friedmann equation in the form $H^2 = {\rho\over 3 M_p^2}
{\cal F}(\rho)$.  From fig.\ \ref{fig3} one notices that the
derivative ${\cal F}\,'(\rho_c)$ diverges at the critical point, even
though ${\cal F}(\rho_c)$ is finite.  If inflation takes place on the
3-brane, starting with $\rho$ close to $\rho_c$, this can give rise to
significant changes in the predictions for the scalar and tensor
spectral indices,
\be
 n_s -1 = {1\over {\cal F}}\left(2\eta - 6\epsilon\left(
	1 + {d\ln{\cal F}\over d\ln\rho}\right)\right), \qquad
 n_t = -{2\epsilon\over{\cal F}}\left(
	1 + {d\ln{\cal F}\over d\ln\rho}\right)
\ee
 and tensor-to-scalar ratio:
\be
r_t = {16\epsilon\over {\cal F}}
\ee
These revert to the standard formulae when ${\cal F}=1$.  To
illustrate the new effect, consider chaotic inflation on the 3-brane with
an $m^2\phi^2$ potential.  We can trade the maximum energy density
parameter $\rho_c$ in the braneworld model for a maximum number of
$e$-foldings of inflation, $N_m$.  As $N_m\to\infty$, we recover the
standard scenario, while if $N_m$ is close to 50 or 60, the effects
will be visible in the CMB spectrum.  Fig.\ \ref{fig4} shows how
the braneworld predictions for $r$ and $n_\sigma$ deviate from the
usual ones, depending on the number of $e$-foldings, which is related to
the scale of inflation.  The modified predictions can be measurably
different from the standard ones, while still being equally good fits
to the present data.  

\begin{figure}[h]
\centerline{\includegraphics[width=0.6\hsize,angle=0]{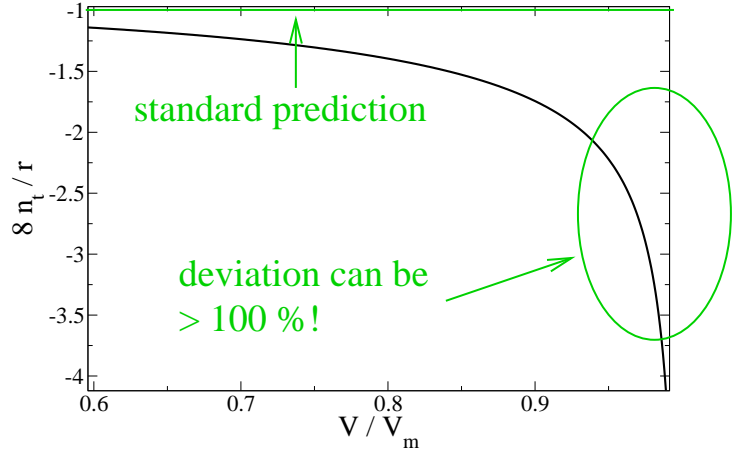}}
\caption{$8n_t/r$ versus $V/V_m$, showing deviation of braneworld
prediction from the standard value, $-1$.
\label{fig5}}
\end{figure}

Furthermore, the deviation of the standard single-field inflation
consistency condition ${n_t/r}=-\frac18$ of can be large; this is
shown in figure \ref{fig5}.  The magnitude depends on how close the
value of the potential $V$ is to the maximum allowed value $V_m$
at horizon crossing of the relevant modes.  Even though $n_t$ will
be very difficult to measure with any accuracy, assuming that the
tensor contribution $r$ will be observed, the discrepancy from the
standard value for $V\cong V_m$ can be a factor of 3 or 4 without much
fine tuning; in this case Planck should be able to detect it, despite
the difficulty of measuring $n_t$ accurately.

\section{Nonlocal inflation and nongaussianity}

It is well known that string theory predicts higher-derivative
corrections to the effective action, including arbitrarily high
orders in derivatives; these are known as $\alpha'$ corrections.
The usual procedure in string cosmology is to work in a low-energy 
regime where these corrections are small enough to safely ignore.
However it is interesting to consider what can be said in the opposite
regime.  Normally this cannot be done in any reliable way because we
do not know the $\alpha'$ corrections to all orders.  However,
$p$-adic string theory provides an exception;\cite{witten} in this theory the
complete action is known to all orders,
\be
	S = {m_s^4\over g_s^2}{p^2\over p-1}\int d^{\,26} x
	\left(-\frac12\phi p^{-\Box/2 m_s^2}\phi + {\phi^{p+1}\over
p+1}\right)
\ee
where $m_s$ is the string scale, $g_s$ the string coupling, and
$p$ = any prime number.  The presence of derivatives at all orders
makes it a nonlocal theory.

\begin{figure}[h]
\centerline{\includegraphics[width=0.5\hsize,angle=0]{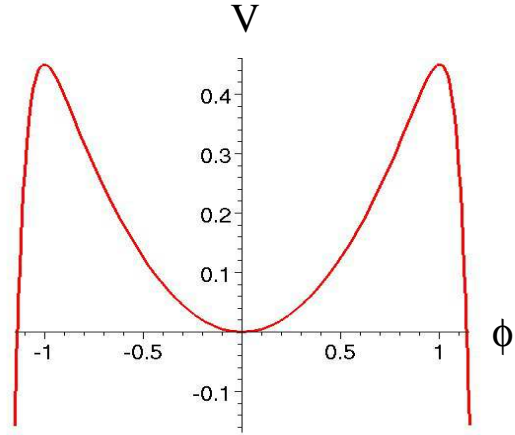}}
\caption{The potential of the tachyon of $p$-adic string theory.
\label{fig6}}
\end{figure}

The field $\phi$ is tachyonic at the maximum
of its potential (fig.\ \ref{fig6}), $\phi=1$, where the curvature would normally be
too great to support inflation.  However in ref.\ \mycite{BBC} we
found the surprising result that inflation from the maximum to
the metastable minimum at $\phi=0$ can indeed occur in the regime
where $\Box/m_s^2$ is not small.  The Hubble scale $H$ is somewhat
above $m_s$ for these solutions, which would normally mean that higher
order corrections are out of control, but in the present case we have
kept them to all orders.

\subsection{Homogeneous solution}

\begin{figure}[h]
\centerline{\includegraphics[width=0.5\hsize,angle=0]{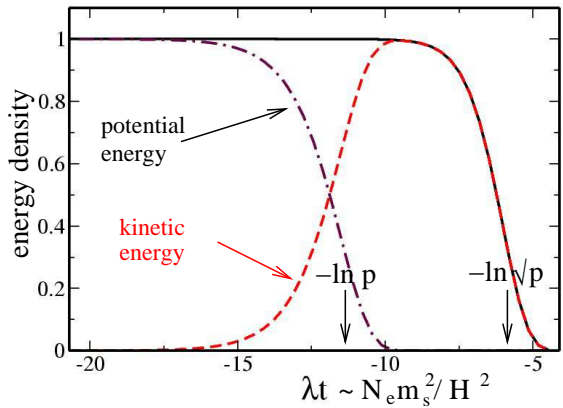}}
\caption{Energy density of $\phi$ versus time in $p$-adic inflation
model, showing how the time of transition between potential and
kinetic energy domination, and the time at which inflation ends, depend
on $p$.
\label{fig7}}
\end{figure}

The Klein-Gordon equation and Friedmann equation are given by 
\be
	e^{-\Box/m_p^2}\phi = \phi^p,\qquad  H^2 = {T_{00}\over 3
M_p^2}
\ee
where  $m_p^2\equiv {2m_s^2\over \ln p}$, and 
the energy density is given by a complicated expression
\be
\rho_{\phi}=-T_{00}={m_s^4\over 2g_p^2}\LT \phi e^{-{\Box\over
m_p^2}}\phi-{2\over p+1}\phi^{p+1}+{1\over m_p^2}\int_0^1 d\tau\ \LF
\Box e^{-{\tau\Box\over m_p^2}}\phi\RF\LF e^{-{(1-\tau)\Box\over
m_p^2}}\phi\RF\right. \nonumber
\ee
\be
\left.+{1\over m_p^2}\int_0^1 d\tau\ \partial_t\LF e^{-{\tau\Box\over
m_p^2}}\phi\RF\partial_t\LF e^{-{(1-\tau)\Box\over m_p^2}}\phi\RF\RT
\label{rho} 
\ee
with ${1\over g_p^2}\equiv {1\over g_s^2}{p^2\over p-1}$. 
Nevertheless we can find approximate solutions when $\phi$ is near
the top of the potential.  When $H>m_s$, we can use the same
friction-dominated approximation for the Laplacian as in ordinary
inflation, $\Box = \partial^2_t + 3 H \partial_t \cong 3H_0
\partial_t$.  We find a solution of the form
\be
\phi(t) =  e^{-e^{\lambda t}},\quad \lambda \cong {2m_s^2\over 3 H_0}
\ee
for which the energy density simplifies,
\be
T_{00}(t) = {m_s^4\over 2g_s^2}\,{p^2\over p-1}\left(
	{p-1\over p+1} e^{-u(p+1)} +\int_u^{up}d\omega\,
	e^{-\omega-pu^2/\omega}\right)
\ee
The first term is potential and the second is kinetic energy.  A
striking difference between this and conventional inflationary
solutions is the fact that slow roll of $\phi$ can occur while
its energy is mainly kinetic; this is the case during the final 
stage of inflation.  Fig.\ \ref{fig7} shows the two contributions
to $T_{00}$ and their sum as a function of time.  For large values
of $p$, the final, kinetic-dominated stage of inflation, can encompass
all the observable $e$-foldings.

\subsection{Fluctuations} For the spectrum of perturbations, we need
to determine the fluctuations, $\phi \ =\  \phi_0(t) + \delta\phi(t,\vec
x)$.  They obey the perturbed Klein-Gordon (KG) equation, 
$p^{-\Box/2m_s^2}\delta\phi = p\delta\phi$.  This may at first look
daunting, but it is easy to see that solutions of the usual KG
equation $-\Box \delta\phi = \omega^2\delta\phi$ also satisfy the
nonlocal one, provided that $\omega^2=2m_s^2$.  However, $\delta\phi$
does not have a canonical kinetic term: it is of the form
$-\frac12 \delta\phi F(\Box) \delta\phi$.   One can do a field
redefinition $\varphi = A\delta\phi \equiv 
(F(\Box)/\omega^2)^{1/2} \delta\phi$ such that $\varphi$ does 
have a standard kinetic term (although its interactions will now
become nonlocal).  Using this field, one can assume the usual relation
between the curvature perturbation and the inflaton fluctuation.  It
leads to the spectral index
\be
n_s -1 \ = \ 3 - 2\nu \ \cong\ 
 -\frac43 \left(m_s\over H\right)^2
\ \cong \ {8 g_s^2 \over p} \left(M_p\over m_s\right)^2
\ee
which shows the need for $H\gg m_s$ and $p\gg g_s^2$ to get
$n_s=0.96$.  The COBE normalization, $A_\zeta = 5\times 10^{-5}$,
puts a further constraint on $g_s^2/p$, 
\be
	{g_s\over\sqrt{p}} = \sqrt{8\over 3\pi^2}\, A_\zeta\,
	{e^{N_e|n_s-1|/2}\over |n_s-1|^{3/2}}\cong  2.5\times 10^{-7}
\ee
which depends on the observable number of $e$-foldings, $N_e$.   These
constraints together determine the string scale,
\be
	m_s\cong 3\times 10^{-7} M_p
\ee

\subsection{Nongaussianity}

One of the interesting features of the $p$-adic inflation is that for
sufficiently large $p$, it predicts an observable level of
nongaussianity, in contrast to conventional single-field inflation
models.\cite{BCNG}  Heuristically this can be explained by the fact that the
nongaussianity in single field models is suppressed by the slow roll
parameters and by $V'''$; making $V'''(\phi_0)$ large enough to
compensate for the slow-roll suppression at some field value $\phi_0$
would spoil slow roll at nearby field values, before sufficient
inflation could occur.  However in nonlocal inflation, the fact that
slow roll can continue despite a steep potential and kinetic energy
domination gives a loophole.  For large values of $p$, $V'''$ can be
made sufficiently large for observable nongaussianity, yet remain
compatible with 60 $e$-foldings of inflation.

Using the uniform curvature gauge in which tensor fluctuations are 
decoupled, the curvature perturbation is $\zeta = -{(H/
\dot\varphi_0)} \delta\varphi$, where $\varphi_0$ is the classical
background solution.  The bispectrum  is then given by
$	\langle \zeta\zeta\zeta \rangle = - 
	\left({H/\dot\varphi_0}\right)^3 
	\langle \delta\varphi\,\delta\varphi\,\delta\varphi \rangle$.
The three-point function for $\delta\varphi$ is related to the 
cubic term in the potential, $
V_{\rm cubic} = 2^{3/2} g_s m_s \delta\varphi^3$.  Comparing to the
definition of the nonlinearity parameter $f_{NL}$,
\be
 \langle \zeta_{k_1}\zeta_{k_2}\zeta_{k_3}\rangle  = -\frac{3}{10}
	(2\pi)^7  f_{NL} \,A_\zeta^4\, {{\scriptstyle\sum_i}
	 k_i^2\over {\scriptstyle \prod_i} k_i^3}
	\,\delta^{(3)}({\scriptstyle\sum_i} \vec k_i)
\ee
one can deduce that
\be
	f_{NL} \cong {5 N_e\over 24\sqrt{2}}\,{\sqrt{p}\over\ln p}\,
	|n_s-1|^2 e^{-N_e|n_s-1|/2} 
\ee
in the $p$-adic inflation model.  The nongaussianity turns out to be
very nearly of the local form.  Interesting values $30 < f_{NL} < 150$
as suggested by the analysis of ref.\ \mycite{Wandelt} can be obtained
by taking $g_s\cong 0.1-0.25$ and $p\sim 10^{11}-10^{12}$.

\end{document}